\documentclass[nofootinbib,twocolumn,showpacs,preprintnumbers,amsmath,amssymb,superscriptaddress]{revtex4-2}

\usepackage{graphicx}
\usepackage{dcolumn}
\usepackage{bm}
\usepackage{wasysym}

\newcommand{\mnras}{MNRAS}

\newcommand{\mz}{\mathfrak{z}}

\begin{document}

\preprint{APS/123-QED}

\title{The impact of the eccentricity on the collapse of an ellipsoid into a black hole}

\author{A.G. Nikiforov}
\email{anikiforov@inasan.ru}
\affiliation{%
Institute of Astronomy, Russian Academy of Sciences, \\ 119017 Moscow, Russia
}%
\author{A.N. Baushev}
\email{baushev@gmail.com, Corresponding author}
\affiliation{
 Bogoliubov Laboratory of Theoretical Physics, Joint Institute for Nuclear Research, \\
    141980 Dubna, Moscow Region, Russia
}
\author{M.V. Barkov}%
 \email{barkov@inasan.ru, barmv05@gmail.com}
\affiliation{%
Institute of Astronomy, Russian Academy of Sciences, \\ 119017 Moscow, Russia
}%

\date{\today}

\begin{abstract}
We consider the gravitational collapse of a homogeneous pressureless ellipsoid. We have shown that the minimal size $r$ that the ellipsoid can reach during collapse depends on its initial eccentricity $e_0$ as $r\propto e_0^\nu$, where $\nu \approx 15/8$, and this dependence is very universal. We have estimated the parameters (in particular, the initial eccentricity) of a homogeneous pressureless ellipsoid, whereat it collapses directly into a black hole.
\begin{description}
\item[Keywords]
Cosmology, Large-scale structure of the universe, cold dark matter, primordial black holes
\end{description}
\end{abstract}

\maketitle

\section{\label{sec:level1} Introduction}
In this paper we consider the collapse of a homogeneous pressureless non-rotating ellipsoid. It may seem strange that we consider the ellipsoid to be completely non-rotating. However, this assumption is almost always true for cosmological objects, because cosmological perturbations with non-zero angular momentum have only the damping mode \cite{zn2}.

The problem has already been considered by many authors. For instance, \cite{linmestelshu1965} considered the gravitational collapse of a uniform spheroid, and \cite{whitesilk1979} --- the collapse of a uniform ellipsoid in the expanding matter-dominated Universe. The interest is quite natural: the problem has numerous astrophysical applications. 

Let us qualitatively consider the collapse of a homogeneous dusty ellipsoid with semi-axes $a_1, a_2, a_3$ without initial velocities of the particles. A homogeneous ellipsoid will remain a homogeneous ellipsoid during the collapse \cite{whitesilk1979}. Since the gravitational attraction is the strongest along the shortest axis (for example, the $z$-axis), the ellipsoid eccentricity grows with time during the collapse. Finally, the ellipsoid transforms into a flat elliptical pancake perpendicular to the $z$ axis. 

The case of a homogeneous sphere $(a_1= a_2= a_3)$ is exceptional. Then the system collapses into a black hole (this conclusion remains true even in the general theory of relativity and even in the presence of pressure, provided that the pressure is not sufficient to stop the collapse \cite{tolman1934}). Apparently, if an ellipsoid only slightly differs from a sphere (its eccentricity is very small), it should form a black hole as well. The aim of this work is to estimate the maximum initial eccentricity at which the ellipsoid still collapses into a black hole.

Generally speaking, the shape of an ellipsoid depends on two ratios (for instance, $a_1/ a_2$ and $a_1/a_3$), but we can avoid of consideration of the two-dimensional parameter space by the following trick. Generally speaking, $a_1\ne a_2\ne a_3$, that is, there are the shortest and longest axes, while the third has an intermediate value. Thus, we may consider only the two extreme cases: the oblate spheroid\footnote{We remind that a spheroid is an ellipsoid with two equal semi-axes, i.e., a spheroid with the circular symmetry.} ($a_1= a_2> a_3$, the 'pumpkin' case), which eccentricity $e$ can be defined as 
\begin{equation}\label{28b1}
e\equiv \sqrt{1-\frac{a^2_3}{a^2_1}}, 
\end{equation}
and the prolate spheroid ($a_1=a_2<a_3$, the 'melon' case), which eccentricity $e$ can be defined as 
\begin{equation}\label{28b2}
e\equiv \sqrt{1-\frac{a^2_1}{a^2_3}}. 
\end{equation}
The spheroid shape in both cases may be characterized by a single parameter, $e$.

The structure of the paper is the following: we derive the equations describing the system in section~\ref{sec:level2}, we solve them and discuss the black hole formation in section~\ref{sec:level3}, and in section~\ref{sec:level4} we discuss the obtained results.

\section{\label{sec:level2}Calculations}
So we consider the gravitational collapse of homogeneous pressureless non-rotating spheroids. We use the Newtonian gravity approximation and assume that the initial velocities of all the spheroid particles are zero. 

The gravitational potential inside an ellipsoid described by following equation \cite{whitesilk1979, linmestelshu1965} 
\begin{equation}
    \Phi_e = -\pi G \rho_e \sum_{i=1}^3 \alpha_i x_i^2,
\end{equation}
where $G$ is the gravitational constant, $\rho_e$ is the ellipsoid density and the coefficients $\alpha_i$ are 
\begin{equation}\label{28eq2}
    \alpha_i = a_1 a_2 a_3 \int_0^{\infty} \frac{d\lambda}{(a_i^2 + \lambda)\prod_{j=1}^3 (a_j^2 + \lambda)^{1/2}},
\end{equation}
where $a_i$ are the lengths of the semi-axes. 

Since we consider spheroids, which are axially symmetric about the $z$-axis, it is reasonable to switch to cylindrical coordinates ($\omega$, $\phi$, $z$). The gravitational potential of the spheroid can be written as,
\begin{equation}\label{28eq1}
    \Phi_s = -\pi G \rho_s \left(\alpha(e) \omega^2 + \beta(e) z^2 \right), 
\end{equation}
where $\rho_s$ is the current density of the spheroid. Being functions of only the current eccentricity $e(t)$, $\alpha(e)$ and $\beta(e)$ can be derived from equation (\ref{28eq2}).

For the case of a spheroid ($a_{1,2}\equiv a$, $a_3\equiv c$ and $\alpha_{1,2} = \alpha$, $\alpha_3=\beta$), this integral can be calculated analytically, which gives us the following result\footnote{There is a misprint in the equation (21) in \cite{linmestelshu1965}.}:
\begin{eqnarray}
    \alpha(e) = \frac{\sqrt{1-e^2}}{e^3} \left( \arcsin{e} -e\sqrt{1-e^2} \right),\label{28eq3} \\
    \beta(e) = \frac{2\sqrt{1-e^2}}{e^3}\left( \frac{e}{\sqrt{1-e^2}} - \arcsin{e} \right),\label{28eq4}
\end{eqnarray}
for the oblate spheroid, and 
\begin{eqnarray}
    \alpha(e) = \frac{1}{e^2}\left(1 - \frac{1-e^2}{2e}\ln{\frac{1+e}{1-e}}\right), \label{28eq5}\\
    \beta(e) = \frac{2(1-e^2)}{e^3}\left(\frac{1}{2} \ln{\frac{1+e}{1-e}} - e\right),\label{28eq6}
\end{eqnarray}
for the prolate case.

Consider a point of the spheroid with initial coordinates ($\omega_0$, $\phi_0$, $z_0$). We need to calculate the point coordinates ($\omega(t)$, $\phi(t)$, $z(t)$) as functions of time. Due to the fact that the spheroid does not rotate $\phi(t)=\phi_0$. It is convenient to introduce new variables  $R(t)\equiv \omega/\omega_0$, $Z(t)\equiv z/z_0$. A homogeneous spheroid remains a homogeneous spheroid during collapse \cite{whitesilk1979}, and as a result, the functions $R(t)$ and $Z(t)$ are the same for any point in the spheroid. Actually, $Z(t)$ and $R(t)$ are just the compression ratios of the spheroid along and across the $z$ axis, respectively. It is clear from the definition that $Z=R=1$ at $t=0$. Since we assume that the initial velocities of all the spheroid particles are zero, $\dot Z=\dot R=0$ at $t=0$, and we obtain the initial conditions:
\begin{equation}\label{28eq10} 
    R(0) = Z(0) = 1, \quad \dot{R}(0)=\dot{Z}(0)=0
\end{equation}
The spheroid eccentricity depends on time and may be expressed\footnote{There is a misprint in the equation (19a) in \cite{linmestelshu1965}.} through $R(t)$, $Z(t)$, and the initial eccentricity $e_0$:
\begin{equation}\label{28eq8} 
    Z^2=R^2 \frac{(1-e^2)}{(1-e_0^2)}
\end{equation}
in the oblate case, and 
\begin{equation}\label{28eq9} 
    R^2=Z^2 \frac{(1-e^2)}{(1-e_0^2)}
\end{equation}
in the prolate case. The equations of motion can be obtained from the Newton's second law and look like \cite{linmestelshu1965}
\begin{subequations}\label{28eq7}
    \begin{eqnarray}
        \frac{d^2 R}{dt^2}&=&-\frac{3 GM}{2a^3}\frac{\alpha(e)}{\sqrt{1-e^2}}\frac{1}{RZ}, \label{28eq7a}\\
        \frac{d^2 Z}{dt^2}&=&-\frac{3 GM}{2a^3}\frac{\beta(e)}{\sqrt{1-e^2}}\frac{1}{R^2}, \label{28eq7b}
    \end{eqnarray}
\end{subequations}
where $M$ is the spheroid mass, $a$ is the equatorial radius of the spheroid. These equations, together with initial conditions~(\ref{28eq10}), fully define the evolution of the system. 

\section{\label{sec:level3}The black hole formation}
Now we need a criterion of the black hole formation. At some moment $t_{col}$, one of the compression factors, $Z(t)$ or $R(t)$, turns to zero. It is $Z(t)$, if the spheroid is oblate (the 'pumpkin' case), and the spheroid at $t=t_{col}$ transforms into a disc of radius $\omega_0 R(t_{col})$, perpendicular to the $z$ axis. In the opposite case, if the spheroid is prolate (the 'melon' case), $R(t)$ turns to zero at $t=t_{col}$, and the spheroid transforms into a needle of length $2 z_0 Z(t_{col})$ along the $z$ axis. 

Let us denote the minimal value that $R(t)$ can reach during the collapse of the oblate spheroid by $R_{fin}$ and the minimal value that $Z(t)$ can reach during the collapse of the prolate spheroid by $Z_{fin}$. We will use two reasonable estimations for $R_{fin}$ and $Z_{fin}$: the simplest 
\begin{equation}
\label{28b3}
 R_{fin}=R(t_{col}), \quad Z_{fin}=Z(t_{col}), 
\end{equation}
and a more sophisticated one, which we call 'with flyby'. For example, let us consider the oblate case. After $t=t_{col}$, the disc starts to expand in the $z$ direction but still shortens in the $x$ and $y$ directions. This means that the estimation $R_{fin}=R(t_{col})$ overestimates $R_{fin}$. A more careful method of estimating $R_{fin}$ is to expand the solution after $t=t_{col}$. Once $Z(t_{col})=0$ is reached, we reverse the $z$ component of the velocity $v_Z \rightarrow -v_Z$ and solve equations~(\ref{28eq7}) (which we may rewrite through ($\{v_R, R, v_Z, Z\}$):
\begin{subequations}\label{28eq11}
    \begin{eqnarray} 
    \frac{dv_R}{dt}&=&-\frac{3 G M}{2 a^3} \frac{\alpha(t)}{\sqrt{1-e_0^2}} \frac{1}{RZ}, \label{28eq11a}\\
    \frac{dR}{dt}&=&v_R, \label{28eq11b} \\
    \frac{dv_Z}{dt}&=&-\frac{3 G M}{2 a^3} \frac{\beta(t)}{\sqrt{1-e_0^2}} \frac{1}{R^2}, \label{28eq11c} \\
    \frac{dZ}{dt}&=&v_Z. \label{28eq11d} 
\end{eqnarray}
\end{subequations}
Then we find the moment when $R=Z$. The spheroid occupies the smallest possible volume at this moment and therefore has the highest density inside a sphere. We have all the reasons to consider the value of $R(t)$ at this time as a better estimation of $R_{fin}$, which we call '$R_{fin}$ with flyby'. 

If we consider a prolate spheroid, we should find '$Z_{fin}$ with flyby' instead, of course. To do so, we act exactly as in the previous case, but we reverse the $R$ component of the velocity $v_R \rightarrow -v_R$ instead of $v_Z$ and substitute to~(\ref{28eq11}) the equations for $\alpha$ and $\beta$ corresponding to the prolate case (see~(\ref{28eq5}, \ref{28eq6})). Then we find the moment after $t_{col}$ when $R=Z$ and name the value of $Z(t)$ at this moment '$Z_{fin}$ with flyby'.

We use two ways to estimate $R_{fin}$ and $Z_{fin}$ for the following reason. Intuitively, the black hole forms if $\omega_0 R_{fin}$ or $z_0 Z_{fin}$ become of the order of the gravitational radius of the spheroid $r_g=2GM/c^2$, where $c$ is the speed of light. However, determining a precise criterion is an extremely difficult task. Equations~(\ref{28eq7}) are derived in the Newtonian approximation, while when a black hole is formed, the effects of general relativity inevitably become significant. At present, exact solutions of general relativity for a collapsing spheroid are unknown. We assume that a black hole forms if the minimal radius that the spheroid reaches during the collapse ($a R_{fin}$ or $a Z_{fin}$) becomes smaller or equal to the radius $2 r_g$ of the marginally bound orbit around the black hole  \cite{Bardeen1972ApJ}: 
\begin{equation}\label{28eq13}
    a R_{fin}=2r_g= \frac{4 G M}{c^2} \quad \text{or}\quad a Z_{fin}=\frac{4 G M}{c^2}.
\end{equation}

\begin{figure}
    \centering
    \includegraphics[width=1.0\linewidth]{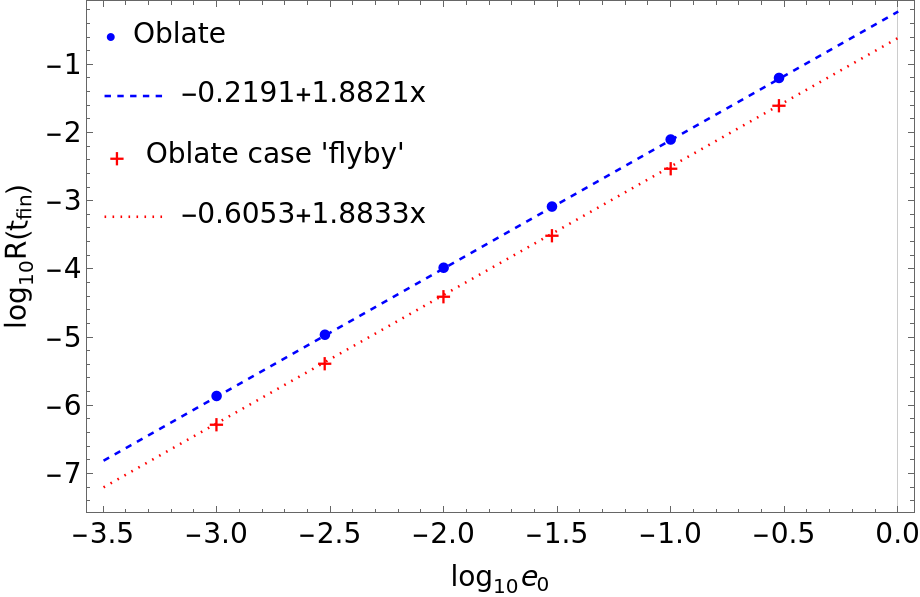}
    \caption{The maximal compression ratio ($R_{fin}$ or $Z_{fin}$), as a function of the initial eccentricity $e_0$ of an oblate spheroid, if we use the simple criterion~(\ref{28b3}) (blue line) and the criterion with flyby (red line).}
    \label{28fig1}
\end{figure}

\begin{figure}
    \centering
    \includegraphics[width=1.0\linewidth]{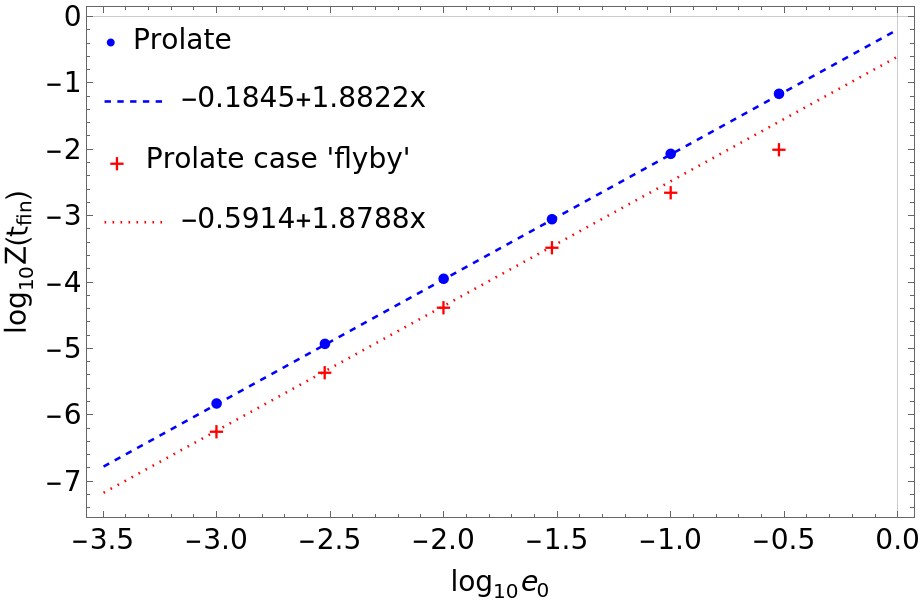}
    \caption{The maximal compression ratio ($R_{fin}$ or $Z_{fin}$), as a function of the initial eccentricity $e_0$ of a prolate spheroid, if we use the simple criterion~(\ref{28b3}) (blue line) and the criterion with flyby (red line).}
    \label{28fig2}
\end{figure}

To find the dependence of $R_{fin}$ or $Z_{fin}$ from the initial eccentricity value, we obtain numerical solutions of~(\ref{28eq11}) for $e_0=\{0.001, 0.003, 0.01, 0.03, 0.1, 0.3\}$, and then fit the obtained dependence $R_{fin}(e_0)$ or $Z_{fin}(e_0)$ by a linear function in the double logarithmic scale. For oblate spheroids, we obtain best fits $\lg R_{fin}=-0.2191 + 1.8821 \lg e_0$ and $\lg R_{fin}=-0.6053 + 1.8833 \lg{e_0}$ for the simple criterion~(\ref{28b3}) and the criterion with flyby, respectively (see Fig.~\ref{28fig1}). For prolate spheroids, we obtain the best fits $\lg Z_{fin} = -0.1845 + 1.8822 \lg e_0$ and $\lg Z_{fin}=-0.5914 + 1.8788 \lg{e_0}$ for the simple criterion~(\ref{28b3}) and the criterion with flyby, respectively (see Fig.~\ref{28fig2}). In the later case, we ignore the last two deviating points at large eccentricities (red dashed line in Fig.~\ref{28fig2}), since we are interested in small values of eccentricity allowing to form a black hole.

\section{\label{sec:level4}Discussion}

Thus, we fit our results with expressions of the form $\mu+\nu \lg e_0$. It is notable that the value of $\nu$ in all four cases turns out to be practically the same $\nu\simeq 1.88\simeq 15/8$. Moreover, the best fits for the oblate and prolate spheroids are also almost the same, if we use the same criterion to determine $R_{fin}$ and $Z_{fin}$. Since $R_{fin}$ and $Z_{fin}$ have the same values if the eccentricity $e_0$ is the same, we may introduce the compression ratio of the spheroid $\chi_{fin}$, which is equal to $R_{fin}$ and $Z_{fin}$ for an oblate and a prolate spheroid, respectively. The best fits are close to 
\begin{equation}\label{28b4}
\lg \chi_{fin}=-0.2 + \frac{15}{8} \lg e_0,
\end{equation}
if we use the simple criterion~(\ref{28b3}), and to 
\begin{equation}\label{28b5}
\lg \chi_{fin}=-0.6 + \frac{15}{8} \lg e_0
\end{equation}
for the criterion with flyby. As we can see, $\mu$ in~(\ref{28b4}) is significantly larger than in~(\ref{28b5}), i.e., $\chi_{fin}$ is $\sim 2.5$ larger if we apply the simple criterion~(\ref{28b3}). It is no wonder: as we discussed in the previous section, criterion~(\ref{28b3}) significantly overestimates $R_{fin}$ and $Z_{fin}$. We will use the with flyby criterion as more accurate.

The fact that the fits for the oblate and prolate spheroids are almost the same is very important: it allows us to generalize our results from spheroids to an arbitrary ellipsoid. In fact, let us consider an ellipsoid with axes $a_1\ge a_2\ge a_3$. We may define its eccentricity $e$ with the help of equation~(\ref{28b1}), and so $e$ does not depend on $a_2$. The value of $a_2$ can range from $a_1$ to $a_3$. Thus, an oblate spheroid with $a_1= a_2\ge a_3$ and a prolate spheroid with $a_1\ge a_2= a_3$ are two limiting cases of all ellipsoids with the same $a_1$ and $a_3$. However, the compression ratio ($R_{fin}$ or $Z_{fin}$) behaves the same in these two cases, and therefore we have all the reasons to suppose that the compression ratio of any ellipsoid with $a_1\ge a_2\ge a_3$ can be well approximated by~(\ref{28b5}), where the eccentricity $e_0$ is given by~(\ref{28b1}). 

Let us apply the obtained results. Hereafter, we assume the standard $\Lambda$CDM cosmology with the present-day radiation fraction $\Omega_{m,0}=5.0\times 10^{-5}$ and $H_0=71$~(km/s) Mpc$^{-1}$ \citep{pdg2022}.  Then the present-day critical density is $\rho_{c,0}\approx 0.97\cdot 10^{-29}$~{g}/cm$^{3}$.

The result we have obtained can be applied to solve many cosmological problems. As an example, let us consider the formation of primordial black holes in the early Universe.
One of the main scenarios~\cite{pasha1994} for their formation is that, for some reason that we will not discuss here, the spectrum of the primary perturbations differs significantly from the Harrison-Zeldovich one \cite{1970PhRvD...1.2726H,1972MNRAS.160P...1Z} in the short-wave region. As a result, the short perturbations become nonlinear already in the early Universe (at the radiation-dominated stage) and collapse forming black holes.

We consider the collapse of a homogeneous ellipsoid with $a_1\ge a_2\ge a_3$. We define the ellipsoid eccentricity $e$ by~(\ref{28b1}). The eccentricity $e$ should be very small for the ellipsoid to be able to collapse into a black hole. Thus, the ellipsoid is initially almost a sphere. We will use this fact. Only  perturbations in dark matter can collapse in the radiation-dominated Universe. Using the method proposed in~\cite{24}, it is easy to show that a matter perturbation stops expanding in the radiation-dominated Universe when its density $\rho_M$ becomes $\sim 7.22$ times higher than the radiation density $\rho_\gamma$
\begin{equation}
\label{28b6}
\rho_M=7.22 \rho_\gamma.
\end{equation}
Since $\rho_\gamma\gg \rho_M$ at the radiation-dominated stage, matter perturbations should be very large to collapse. Suppose that a perturbation of mass $M$ collapses at the redshift $\mz\gg1$. Let us estimate the maximum eccentricity $e_0$ at the beginning of the collapse, which allows us to form a black hole. The initial radius $a$ of the ellipsoid (we remind that it is almost a sphere) is bound with $M$ by,
$$
M=\frac43\pi a^3 \rho_M= 7.22\frac43\pi a^3 \rho_\gamma\approx 30 a^3 \Omega_\gamma \rho_{c0}\mz^4.
$$
Here we use~(\ref{28b6}). So
\begin{equation}
\label{28b7}
a\approx\mz^{-4/3}\left({\frac{M}{30\Omega_\gamma \rho_{c0}}}\right)^{1/3}.
\end{equation}
Substituting this value to~(\ref{28eq13}), we get
\begin{equation}
\label{28b8}
\chi_{fin}\simeq 1.2\times 10^{-4} \mz^{4/3}_9 M^{2/3}_{33.3}.
\end{equation}
Here we have introduced the notations $M_{33.3}$, which is the ellipsoid mass divided by the solar mass $M_{33.3}=M/ 2\times10^{33} \rm g$, and  $\mz_9\equiv\mz/10^9$. Comparing this result with~(\ref{28b5}), we find
\begin{equation}
\label{28b9}
e_0\simeq 1.7\cdot 10^{-2} \mz_9^{32/45} M_{33.3}^{16/45}.
\end{equation}
For instance, if the ellipsoids of mass $1 M_\odot$ collapse in the early Universe at the temperature $\sim 240$~{keV},  they can form black holes if their eccentricities are relatively modest ($e_0\sim 0.02$).
\begin{acknowledgments}
This work was partially supported by the Russian Science Foundation grant 23-22-00385.
\end{acknowledgments}


\providecommand{\noopsort}[1]{}\providecommand{\singleletter}[1]{#1}%

\end{document}